\documentclass[twoside]{article}

\usepackage{amsmath,amsthm,amssymb}

\usepackage{newtxtext,newtxmath}

\usepackage[text={12.5cm,19cm},centering,paperwidth=17cm,paperheight=24cm]{geometry}

\usepackage[fontsize=10.4pt]{scrextend}

\def\titlerunning#1{\gdef\titrun{#1}}

\def\R {\mathbb{R}}                          
\def\e{\rm{e}}
\def\su{$SU(1,1)$}
\def\ax{$AX+B$ }    \def\rc{\rho^{\text{cl}}}      
\def\rc{\rho^{\text{cl}}}
\def\sq{S^\text{Q}}
\def\tr{\text{Tr}}                   
\def\dd{{\, \rm d} }             
\def\C {\mathbb{C}}
\def\D{\mathbb{D}}
\def\cD {\mathcal{D}}                  
\def\C {\mathbb{C}}       
\def\cF {\mathcal{F}}
\def\cH{\mathcal{H}}
\def\sc{S^\text{cl}}                     

\makeatletter
\def\author#1{\gdef\autrun{\def\and{\unskip, }#1}\gdef\@author{#1}}
\def\address#1{{\def\and{\\\hspace*{15.6pt}}\renewcommand{\thefootnote}{}\footnote{#1}}\markboth{\autrun}{\titrun}}
\makeatother

\def\email#1{email: \href{mailto:#1}{#1} }

\newenvironment{dedication}{\itshape\center}{\par\medskip}
\newenvironment{acknowledgments}{\bigskip\small\noindent\textit{Acknowledgments.}}{\par}

\newtheorem{thm}{Theorem}[section]

\newtheorem{prob}[thm]{Problem}
\newtheorem{conjecture}[thm]{Conjecture}

\newtheorem{mainthm}[thm]{Main Theorem}

\theoremstyle{definition}
\newtheorem{defin}[thm]{Definition}

\newtheorem*{rem}{Remark}

\numberwithin{equation}{section}

\usepackage[hyperfootnotes=false,colorlinks=true,allcolors=blue]{hyperref}

\begin{document}

\titlerunning{Coherent state entropy inequalities for $SU(1,1)$ and its $AX+B$ subgroup }

\title{\textbf{Wehrl-type coherent state entropy
inequalities for $SU(1,1)$ and its $AX+B$ subgroup}}

\author{Elliott H. Lieb  \and Jan Philip Solovej}

\date{}

\maketitle

\address{E.H. Lieb:  Departments of Mathematics and Physics, Jadwin Hall,
Princeton University, Princeton, NJ 08544, USA; \email{lieb@princeton.edu} \and J.P. Solovej:  QMATH, Department
  of Mathematical Sciences, University of Copenhagen,
  Universitetsparken 5, DK-2100 Copenhagen \O, Denmark ; 
  \email{solovej@math.ku.dk}}

\begin{dedication}
Dedicated to our friend and colleague   Ari~Laptev on his 70$^{\rm th}$ birthday.
\end{dedication}

\begin{abstract}
 We discuss the the Wehrl-type entropy inequality conjecture for the
 group $SU(1,1)$ and for its subgroup $AX+B$ (or affine group), their
 representations on $L^2(\R_+)$, and their coherent states.  For
 $AX+B$ the Wehrl-type conjecture for $L^p$-norms of these coherent
 states (also known as the R\e nyi entropies) is proved in the case
 that $p$ is an even integer.  We also show how the general $AX+B$
 case reduces to an unsolved problem about analytic functions on the
 upper half plane and the unit disc.
\end{abstract}

\section{Introduction}

\label{intro}
 \setcounter{footnote}{0}
The starting point for this work\footnote{This paper is a polished version of a paper on arxiv \cite{LS4}. It is
not the final version, however, and should be regarded as work in
progress.}, historically, was Wehrl's definition
of semiclassical entropy and his conjecture about it's minimum value
\cite{W}. Given a density matrix $\rho$ on $L^2(\R)$ (a positive
operator whose trace is 1) we define its classical probability density
(or Husimi function) as follows:
\begin{equation}
         \rc (p,q) := \langle p,q | \rho | p,q \rangle.
\end{equation}

Here
$|p,q\rangle$ is the (Schr\"odinger, Klauder, Glauber) coherent state,
which is a normalized function in $L^2(\R)$ parametrized by $p,q \in
\R^2$, which is the classical phase space for a particle in
one-dimension. It is given (with $ \hbar =1$) by
\begin{equation}
|p,q\rangle (x) = \pi^{-\tfrac{1}{4 } }\exp[-(x-q)^2 +ipx].  
\end{equation}
We use the usual Dirac notation in which $|\cdots\rangle$ is a vector
and $\langle\cdots|\cdots\rangle$ is the inner product conjugate
linear in the first variable and linear in the second.

The von Neumann entropy of any quantum state with density matrix $\rho $ is
$$
S^{\text{Q}}(\rho) :=-\tr \rho
\log \rho
$$
and the classical entropy of any continuous probability
density $\widetilde\rho (p,q)$ is
$$S^{\text{cl}}(\widetilde\rho)
:=-\int \widetilde\rho \ \log \widetilde\rho \dd p \dd   q.
$$
The $\sq $
entropy is non-negative, while $\sc(\widetilde\rho) \geq 0$ for any $\widetilde\rho$
that is pointwise $\leq 1$, as is the case for $\rc$.

For the reader's convenience we recall some basic facts of, and the
interest in, the Wehrl entropy.  A classical probability does not
always lead to a positive entropy and it often leads to an entropy
that is $-\infty$. The quantum von Neumann entropy is always
non-negative. Wehrl's contribution was to derive a classical
probability distribution from a quantum state that has several very
desirable features. One is that it is always non-negative. Another is
that it is monotone, meaning that the Wehrl entropy of a quantum state
of a tensor product is always greater than or equal to the entropy of
each subsystem obtained by taking partial traces. This monotonicity
holds for marginals of classical probabilities, but not always for the
quantum von Neumann entropy (the entropy of the universe may be
smaller that the entropy of a peanut).  In short the Wehrl entropy
combines some of the desirable properties of the classical and quantum
entropies.

The minimum
possible von Neumann entropy is zero and occurs when $\rho $ is any
pure state, while that of the classical $\sc(\rc)$ is strictly
positive; Wehrl's conjecture was that the minimum is 1 and occurs when
$$
\rho = |p,q\rangle \langle p,q|. 
$$
That is, when $\rho $ is a pure state projector onto any coherent state.
This conjecture was proved by one of us \cite{L1}. The uniqueness of
this choice of $\rho$ was shown by Carlen \cite{CARLEN}. Recently De
Palma \cite{DEPALMA} determined the minimal Wehrl entropy when the von
Neumann entropy of $\rho$ is fixed to some positive value.

The Wehrl conjecture was generalized in \cite{L1} to the theorem that
the operator norm $\Vert \rc \Vert_p $, $ p>1$ was maximized for the
same choice of $\rho$. This is equivalent to saying that the classical
R\e nyi entropy of $\rc$ is minimized for this same choice of $\rho$.
Later, in our joint paper \cite{LS1}, the same upper bound result was
shown to hold for the integral of \textit{any} convex function of
$\rc$, not just $x\to x^p$. Note that minus a convex function is a
concave function so maximizing integrals of convex functions is the
same as minimizing integrals of concave functions.

We achieved the generalization to all convex functions in \cite{LS1}
by considering a generalization of the map from quantum states to the
classical Husimi functions.  Instead of only considering maps from
quantum states to classical states we may consider maps between
quantum states. In 1991 in \cite{LS3} we defined a general formalism for defining
such maps between different unitary representations of the same group. In \cite{LS3} we named these maps {\it quantum coherent
  operators}. They are, in fact, examples of quantum channels, i.e.,
completely positive trace preserving maps. Moreover, they are
covariant in the sense that applying a unitary transformation to the
state prior to the map is the same as applying it after the map.
Later such covariant quantum channels were studied in
\cite{nuwairan13,nuwairan14,brannan18,brannan20,holevo}.

The possibility of using coherent states to associate a classical
density to a quantum state $\rho$ has a group-theoretic
interpretation, which we will explain below. In the case of the
classical coherent states discussed above this relates to the
Heisenberg group and the coherent states are heighest weight vectors
in the unitary irreducible representation.  This suggests that we can
look at other groups, their representations, and associated coherent
states (heighest weight vectors) and ask whether the analog of the
Wehrl conjecture holds. The first quesiton in this direction was
raised in \cite{L1} for the group $SU(2)$ where the representations
are labeled by the quantum spin $J=0,1/2,1,3/2,\ldots$. The spin cases $J=1/2, 3/2$
were proved by Schupp in \cite{SCHUPP} and in full generality by
us in \cite{LS1}.

The obvious next case is $SU(N)$ for general $N$ which has many kinds
of representations.  We showed the Wehrl hypothesis for all the
symmetric representations in \cite{LS2}. Here we turn our attention to
another group of some physical but also mathematical interest which is
the group $SU(1,1)$ and its subgroup the \ax group or affine group. These groups,
like the Heisenberg group, are not compact and have only infinite
dimensional unitary representations.  The affine group is not even
unimodular which means that the left and right invariant Haar measures
are different. The group $SU(1,1)$ is not simply connected. The purely
mathematical interest in the affine group is in signal processing (see \cite{Dau}).
The coherent states for the affine group are like continuous families of
wavelets.

The \ax (or affine) group  is $G=\R_+\times\R$ with the composition rule
$$
(a,b)\cdot(a',b')=(aa',ab'+b).
$$
This comes from thinking of the group acting on the real line as
$$
x\mapsto ax+b,
$$
hence the name  $ax+b$ group.
It is \textit{not} unimodular and
the {\it left} Haar measure on the group is given by
$$
a^{-2}\dd  a\dd  b ,
$$
(for reference, the right Haar measure is  $a^{-1} \dd a \dd b$.)

The group $SU(1,1)$ is the group of $2\times2$ complex matrices of determinant one, that leave the
form $|z_1|^2-|z_2|^2$ invariant, i.e., all matrices of the form
\begin{equation}\label{eq:su11matrix}
  \left(\begin{array}{cc} \alpha&\beta\\\overline\beta&\overline\alpha 
  \end{array}\right),
\end{equation}
where $|\alpha|^2-|\beta|^2=1$. It is easily seen that we may consider the affine \ax group as a subgroup
of $SU(1,1)$ through the injective group homomorphism
$$
(a,b)\mapsto(\alpha,\beta)=\frac1{2\sqrt{a}}(a+1+ib, a-1+ib).
$$
Both the \ax group and $SU(1,1)$ act transitively on the open complex unit disk by
$$
   z\mapsto \frac{\alpha z+\overline\beta}{\beta z+\overline\alpha}.
$$
This is equivalent to a transitive action on the upper half-plane. For the \ax group
the action on the upper half plane is particularly simple as it becomes
$$
z\mapsto az+b.
$$
The coherent states for the two groups will naturally be parametrized
by points on the unit disk or equivalently by points on
the upper half plane.

In Section~\ref{sec:axbrep} we introduce the unitary representations
of the affine group and its coherent states and state the Wehrl
entropy conjecture in this case.  In Section~\ref{sec:genwehrl}  we generalize the
conjecture to a larger class of functions, not just the entropy and we
give an elementary proof in special cases.  In Section~\ref{sec:analyticform} we
formulate the generalized Wehrl inequality as a statement about
analytic functions on the upper half plane or the unit disk. In
Section~\ref{sec:su(1,1)} we consider $SU(1,1)$ and a
discrete series of unitary representations and its corresponding
coherent states.  We formulate the generalized Wehrl conjecture for
$SU(1,1)$ and show that it is a consequence of the generalized Wehrl
conjecture for the affine \ax group.  In Section~\ref{sec:su(1.1)channels} we define for
$SU(1,1)$ certain covariant quantum channels that map density matrices
on one representation space to density matrices on another. These
quantum channels are analogous to those $SU(2)$ and $SU(N)$ channels
we introduced in \cite{LS1,LS2}. We conjecture that the majorization
results from \cite{LS1, LS2} hold also in the case of $SU(1,1)$. We
finally prove that this majorization conjecture implies our
generalized Wehrl inequality by taking an appropriate semi-classical
limit.


\section{Unitary representations and coherent states for the $AX+B$ group}
\label{sec:axbrep}

We begin with the subgroup of $SU(1,1)$ since that is easier than the full
$SU(1,1)$ group and where we have the most
results.

The group has two irreducible faithful unitary representations \cite{AK,GN}
which
may be realized on $L^2(\R_+,dk)$ either by
$$
[U(a,b)f](k)=\exp(-2\pi ibk)a^{1/2}f(ak)
$$
or with $e^{-2\pi ibk}$ replaced by $e^{+2\pi ibk}$. For the representation above the coherent states
we shall consider are given in terms of fiducial vectors
$$
\eta_{\alpha}(k)= C(\alpha)k^\alpha\exp(-k)
$$
where the parameter $\alpha$ is positive and which we will henceforth keep
fixed.  These functions are identified as affine coherent states
(extremal weight vectors) for the representation of the affine group
in \cite{DauKlauPaul,KlauderWatson}.  The constant $C(\alpha)$ is
chosen such that $\int_0^\infty |\eta_{\alpha}(k)|^2/k\dd k=1$, i.e.,
$$
C(\alpha)= 2^\alpha\,{\Gamma(2\alpha)}^{-1/2}.
$$
Then
\begin{equation}\label{eq:resolution}
  \int_{-\infty}^\infty \int_0^\infty
  U(a,b)|\eta_{\alpha}\rangle\langle\eta_{\alpha}|U(a,b)^* a^{-2}\dd  a\dd b=I.
\end{equation}
For any function $f\in L^2(\R_+)$ we may introduce the {\bf coherent state transform}
\begin{equation}\label{eq:transform}
  h_f(a,b)=\langle U(a,b)\eta_{\alpha}|f\rangle=C(\alpha)
  a^{\alpha+1/2} \int_0^\infty\exp(- k a+2\pi ibk)k^\alpha f(k) \dd k.
\end{equation}
As in the case of the classical coherent states, we then have
$$
\int_{-\infty}^\infty \int_0^\infty |h_f(a,b)|^2 a^{-2}\dd a\dd b=\int_0^\infty|f(k)|^2\dd k,
$$
and if $f$ is normalized in $L^2$ we may consider $|h_f(a,b)|^2$ as a probability density relative to the
Haar measure $a^{-2}\dd a\dd b$. Observe that in contrast to the classical case we do not have
$|h_f(a,b)|^2\leq 1$, but rather
$$|h_f(a,b)|^2\leq \int|\eta_\alpha(k)|^2  \dd k=\alpha.$$
This is a consequence of the fact
that \eqref{eq:resolution} requires $\int_0^\infty |\eta_{\alpha}(k)|^2/k\dd k=1$, which is different from the
$L^2$-normalization. This difference is due to the group not being uni-modular.
The {\bf Wehrl entropy}
\begin{equation}
  S^{\rm W}(f)= -\int |h_f(a,b)|^2\ln(|h_f(a,b)|^2) a^{-2} \dd a\dd  b  
\end{equation} 
is, therefore, {\it not necessarily non-negative}, but by the definition,
and by the normalization of $h_f$, it is bounded below by $-\ln
\alpha$.  The natural generalization of Wehrl's conjecture in this
case is that the entropy $S^{\rm W}(f)$ is minimal for $f= \alpha^{-1/2} \eta_\alpha$, in which case it is
$1+ (2\alpha)^{-1} -\ln \alpha$.

In the following section we state a more general conjecture and prove
it in special cases.  In the last section of the paper we show that our
conjecture and  theorem are equivalent to $L^p$ estimates for
analytic functions in the complex upper half plane.



\section{Generalized conjecture for the \ax group and a partial result}\label{sec:genwehrl}
We make the following more general conjecture.

\begin{conjecture}[Wehrl-type conjecture for the \ax group]
  \label{conj:1} Let $\alpha>0$ be fixed. If $G:[0,\alpha]\to \R$ is a convex function then
$$
  \int_{-\infty}^\infty \int_0^\infty  G(|h_f(a,b)|^2)a^{-2}\dd a\dd b 
$$
  is maximized among all normalized $f\in L^2(\R_+)$ if and only if (up to a phase) $f$ has the form
$f=\alpha^{-1/2}U(a,b)\eta_{\alpha}$ for some $a>0$ and $b\in\R$.
\end{conjecture}

\begin{rem} {\rm The maximal value above may be infinite.
If $G(t)=t^{s}$ for $s\geq 1$ the maximal value of the integral is conjectured to be
$$
  \frac{2\alpha^s}{(2\alpha+1)s-1}.
$$
The analog of Wehrl's original entropy conjecture follows from this
conjecture by taking minus a derivative at $s=1$ as in \cite{L1} and gives
the minimal entropy
\begin{equation}
  S^W_{\rm min} = 1+(2\alpha)^{-1}-\ln\alpha,
\end{equation}
which agrees with the lower
bound $-\ln\alpha$ mentioned above.  }
\end{rem}

For $G(t)=t^{s}$ with $s=1+(2\alpha+1)^{-1}$ the conjecture was proved in \cite{BAND} by J.~Bandyopadhyay.
The following theorem, which has an elementary proof, yields the conjecture for integer $s$.
As we explain in the next section this result also follows
from an application of Theorem~3.1 of \cite{BURBEA} and from \cite{BBHOP} in their works on holomorphic functions.
An alternative proof with a few generalizations of the result below was given recently in \cite{bekole}.
In \cite{BBHOP} a conjecture is formulated
that is equivalent to the above for $G(t)=t^{s}$ for all $s\geq1$.
\begin{mainthm}
 The statement of Conjecture~\ref{conj:1} holds for the special cases
of $G(t)=t^{s}$, for $s$ being a positive integer.
\end{mainthm}

\begin{proof}
We want to prove that
$$
\int_{-\infty}^\infty \int_0^\infty  |h_f(a,b)|^{2s}a^{-2}\dd a\dd b 
$$
is  maximized if and only if
$$
f(k)= Ak^\alpha\exp(-Bk)
$$
with $A,B\in \C$ such that $f\in L^2(\R_+)$ is normalized.
The proof will rely only on a Schwarz inequality and existence and uniqueness will follow from the corresponding
uniqueness of optimizers for Schwarz inequalities.

If $s$ is a positive integer we can write

\begin{align}
 h_f(a,b)^s=C(\alpha)^sa^{s(\alpha+1/2)}&\int_0^\infty\cdots\int_0^\infty
  \exp\left[(-a+2\pi i b)(k_1+\cdots+k_s)\right]\nonumber\\
    &\times(k_1\cdots k_s)^\alpha f(k_1)\cdots f(k_s)\dd k_1\cdots
  \dd k_s.\nonumber
\end{align}
Hence doing the $b$ integration gives

\begin{align}
  \int_{-\infty}^\infty\int_0^\infty & |h_f(a,b)|^{2s}a^{-2}\dd a \dd b \\= &C(\alpha)^{2s}\Gamma(s(2\alpha+1)-1)\int_0^\infty\cdots\int_0^\infty
      [2(k_1+\cdots k_{s})]^{-s(2\alpha+1)+1}\nonumber\\
      &\times k_1^\alpha f(k_1)\cdots k_s^\alpha f(k_s)k_{s+1}^\alpha \overline{f(k_{s+1})}\cdots k_{2s}^\alpha\overline{f(k_{2s})}
      \dd k_1\cdots\dd k_{2s-1}.\nonumber
\end{align}

where $k_{2s}=k_1+\cdots+k_s-(k_{s+1}+\cdots +k_{2s-1})$.

We now do the $a$ integration and arrive at

\begin{align}
  \int_{-\infty}^\infty\int_0^\infty &|h_f(a,b)|^{2s}a^{-2}\dd a \dd b \\= &C(\alpha)^{2s}\Gamma(s(2\alpha+1)-1)\int_0^\infty\cdots\int_0^\infty
      [2(k_1+\cdots k_{s})]^{-s(2\alpha+1)+1} \nonumber\\
      &\times k_1^\alpha f(k_1)\cdots k_s^\alpha f(k_s)k_{s+1}^\alpha \overline{f(k_{s+1})}\cdots k_{2s}^\alpha\overline{f(k_{2s})}
      \dd k_1 \cdots \dd k_{2s-1}.  \nonumber
\end{align}

We now change variables to
\begin{align}
 r&=k_1+\cdots +k_s\nonumber\\
  u_j&=k_j/r,&&j=1,\ldots,s-1\\
  v_j&=k_{s+j}/r,&&j=1,\ldots,s-1.\nonumber
\end{align}

The Jacobian determinant for this change of variables is easily found to be
$$
  \left|\text{det}\left[\frac{\partial(k_1,\ldots,k_{2s-1})}{\partial(r,u_1,\ldots,u_{2s-1},v_1,\ldots,v_{2s-1})}\right]\right|
  =r^{2s-2}.
$$
We arrive at
 \begin{align}
   \int_{-\infty}^\infty\int_0^\infty &|h_f(a,b)|^{2s}a^{-2}dadb\nonumber \\= &C(\alpha)^{2s}\Gamma(s(2\alpha+1)-1)
   \int_0^\infty
   \idotsint\displaylimits_{u_1+\cdots+u_{s-1}\leq1}\,\,\,\,\idotsint\displaylimits_{v_1+\cdots+v_{s-1}\leq1}
   2^{-s(2\alpha+1)+1} r^{s-1}\nonumber\\&\times [u_1\cdots u_{s-1}(1-u_1-\cdots-u_{s-1})]^\alpha
   \\&\times[v_1\cdots v_{s-1}(1-v_1-\cdots-v_{s-1})]^\alpha
   \nonumber\\&\times f(u_1r)\cdots f(u_{s-1}r)f( r(1-u_1-\cdots-u_{s-1}))\nonumber\\&\times
    \overline{f(v_1r)}\cdots \overline{f(v_{s-1}r)}\overline{f( r(1-v_1-\cdots-v_{s-1}))}\nonumber\\&\times
    \dd u_1\cdots\dd u_{s-1}\dd v_1\cdots\dd v_{s-1}\dd r.\nonumber
 \end{align}
Let us apply the Cauchy-Schwarz inequality for each fixed $r$ to conclude that
\begin{align}
     &\int_{-\infty}^\infty\int_0^\infty
     |h_f(a,b)|^{2s}a^{-2}\dd a \dd b\nonumber
     \\&\leq 2^{-s(2\alpha+1)+1}C(\alpha)^{2s}\Gamma(s(2\alpha+1)-1)
     \nonumber\\&\times\biggl[\idotsint\displaylimits_{u_1+\cdots+u_{s-1}\leq1}
       [u_1\cdots u_{s-1}(1-u_1-\cdots-u_{s-1})]^{2\alpha}\dd
       u_1\cdots\dd u_{s-1}\biggr]\nonumber\\&
     \times\int_0^\infty\biggl[\idotsint\displaylimits_{u_1+\cdots+u_{s-1}\leq1}|f(u_1r)\cdots
       f(u_{s-1}r)f( r(1-u_1-\cdots-u_{s-1}))|^2 \nonumber\\&\qquad\times\dd
       u_1\cdots\dd u_{s-1}\biggr] r^{s-1}\dd
     r\nonumber\\ &=2^{-s(2\alpha+1)+1}C(\alpha)^{2s}\Gamma(s(2\alpha+1)-1)\left(\int_0^\infty|f(r)|^2\dd
     r\right)^s\\&\times
     \biggl[\idotsint\displaylimits_{u_1+\cdots+u_{s-1}\leq1}
       [u_1\cdots u_{s-1}(1-u_1-\cdots-u_{s-1})]^{2\alpha}\dd
       u_1\cdots\dd u_{s-1}\biggr].\nonumber
  \end{align}
  
If $f$ is normalized this is a number depending on $\alpha$ and\
$s$. The important observation is that the upper bound is achieved if
and only if there is a function $K(r)$ depending on $r$ such that
$$
f(u_1)\cdots f(u_{s-1})f( r-u_1-\cdots-u_{s-1})
=K(r)[u_1\cdots u_{s-1}(r-u_1-\cdots-u_{s-1})]^\alpha
$$
for almost all $0\leq u_1,\ldots,u_{s-1},r$ satisfying
$u_1+\cdots+u_{s-1}\leq r$.
If we define $g(u)=u^{-\alpha}f(u)$ and introduce the variable $u_s=r-(u_1+\cdots+u_{s-1})$ we may rewrite this is as
$$
g(u_1)\cdots g(u_s)=K(u_1+\cdots+u_s)
$$
for almost all $0\leq u_1,\ldots,u_s$. It is not difficult to show
that if a locally integrable function $g$ satisfies this it must be smooth and it
follows easily that
$$
g(u)=A\exp(-Bu)
$$
for complex numbers $A,B$ which is exactly what we wanted to prove.

The maximal value can be found by a straightforward computation.
\end{proof}


\section{An analytic formulation}\label{sec:analyticform}
Using the Bergman-Paley-Wiener Theorem in \cite{Durenetal}
we can rephrase our conjecture and
theorem in terms of analytic functions on the complex upper half plane $\C_+=\{z\in \C\ |\ \Im z>0\}$.
Introducing the weighted Bergman space,
$$
{\mathcal A}_{\beta}^2(\C_+)=\{F\in L^2(\C_+,(\Im z)^\beta d^2 z) \ |\ F \text{ analytic } \}
$$
for $\beta>-1$.
The Paley-Wiener Theorem in this context says that there is a unitary map
$$
L^2(\R_+)\ni f\mapsto F\in {\mathcal A}_{2\alpha-1}^2(\C_+),
$$
given by
$$
F(z)= \frac{2^\alpha}{\sqrt{2\pi\Gamma(2\alpha)}} \int_0^\infty e^{ikz}k^{\alpha}f(k)dk. 
$$
If we write $z=2\pi b+ia$ we recognize the coherent state transform in \eqref{eq:transform} to be
$$
h_f(a,b)=\sqrt{2\pi}(\Im z)^{\alpha+1/2} F(z). 
$$
The action of the
affine group on $\R$ may be extended to the upper half plane $\C_+$ by
$z\mapsto az+2\pi b$. The representation of the \ax group on the Bergman space is then
$$
 (U(a,b)^* F)(z)=\frac{2^\alpha}{\sqrt{2\pi\Gamma(2\alpha)}}\int_0^\infty e^{ikz}k^{\alpha}(U(a,b)^*f)(k)\dd k
  =a^{\alpha+1/2}F(az+2\pi b)
$$
In the analytic representation our conjecture states that
if $G:[0,\alpha/(2\pi)]\to\R$ is convex then
\begin{equation}\label{eq:analyticHP}
\int_{\C_+} G(|F(z)|^2(\Im z)^{2\alpha+1})|\Im z|^{-2} \dd^2 z
\end{equation}
is maximized among normalized functions $F$ in ${\mathcal A}^2_{2\alpha-1}(\C_+)$ if and only if
$F$ is proportional to $(z-z_0)^{-(2\alpha+1)}$ for some $z_0$ in
the lower half-plane.

We may also formulate the conjecture for analytic functions on the unit disk $\D$. Here it states that
if $G:[0,\alpha/(2\pi)]\to \R$ is convex then
\begin{equation}\label{eq:analyticUD}
  \int_{\D} G(|F(z)|^2(1-|z|^2)^{2\alpha+1})(1-|z|^2)^{-2}\dd^2z
\end{equation}
is maximized among all analytic functions $F$ on the unit disk
with
$$
\int_\D|F(z)|^2 (1-|z|^2)^{2\alpha-1}   \dd^2    z=1
$$
if and only if $F(z)$ is proportional to $(1-\zeta z)^{-(2\alpha+1)}$ for some $\zeta$ inside the unit disk.

Our theorem from the previous section is therefore equivalent to the following result.
\begin{thm} If $G(t)=t^s$, with $s$ a positive integer,
then the integrals \eqref{eq:analyticHP} and \eqref{eq:analyticUD}
satisfy the maximization properties stated above.
\end{thm}

\section{Some unitary $SU(1,1)$ representations and their coherent states}
\label{sec:su(1,1)}
We may represent $SU(1,1)$ unitarily on the bosonic Fock space over two modes
$$
\cF(\C^2)=\bigoplus_{N=0}^\infty\cH_N,\qquad\cH_N=\otimes_{\text{Sym}}^N\C^2,\qquad \cH_0=\C.
$$ On this space we have the action of the creation and annihilation
operators $a_1,a_2$ and $a_1^\dagger,a_2\dagger$ annihilation and
creation particles in the two modes represented by the standard basis
in $\C^2$.  The unitary representation of $SU(1,1)$ on $\cF(\C^2)$ is
such that the element \eqref{eq:su11matrix} in $SU(1,1)$ acts as the
unitary that transforms the creation and annihilation operators
according to
$$
\left(\begin{array}{c} a_1\\a_2^\dagger
  \end{array}\right)\rightarrow\left(\begin{array}{cc} \alpha&\beta\\\overline\beta&\overline\alpha 
  \end{array}\right)\left(\begin{array}{c} a_1\\a_2^\dagger
  \end{array}\right)=\left(\begin{array}{c} a_1+\beta a_2^\dagger\\\overline\beta a_1+\overline a_2^\dagger
  \end{array}\right),
$$
i.e., a Bogolubov transformation.
It is is clear that this defines a unitary representation of $SU(1,1)$. It is however not irreducible on $\cF(\C^2)$.
We see that the operator
$$
\widehat K=a_1^\dagger a_1-a_2^\dagger a_2
$$
is left invariant by the action of the Bogolubov transformation. Hence we may write
$$
\cF(\C^2)=\bigoplus_{K=-\infty}^\infty\cD_K,
$$
where $\cD_K$ is the subspace where $\widehat K$ is equal to
$K$. Each of the spaces $\cD_K$ are invariant and irreducible for the action of
$SU(1,1)$ moreover the representation on $\cD_K$ is the same as the
representation $\cD_{-K}$ as this corresponds just to the interchanging $a_1$, $a_2$. We will thus focus on $\cD_K$ for
$K\geq0$.

The Lie algebra of $SU(1,1)$ is generated by the three elements
$\widehat K_0$, $\widehat K_\pm$ with the commutation relations (see \cite{PERELOMOV})
$$
[\widehat K_0,\widehat K_\pm]=\pm \widehat K_\pm,\quad [\widehat K_-,\widehat K_+]=2\widehat K_0.
$$
On the Fock space they are represented as
 $$
\widehat K_0=\frac12(a_1^\dagger a_1+a_2^\dagger a_2+1),\quad \widehat K_+=a_1^\dagger a_2^\dagger,\quad \widehat K_-=a_1a_2.
$$
We see that an orthonormal basis for $\cD_K$,$K\geq0$ is given by
$$
|n\rangle_K=\left(\frac{n!(n+K)!}{K!}\right)^{1/2}K_+^n|0\rangle_K,
$$
where $|0\rangle_K$ is defined (up to a phase) by
$$
a_1^\dagger a_1|0\rangle_K=K|0\rangle_K,\quad a_2|0\rangle_K=0.
$$
This easily shows that $\cD_K$ define irreducible representation spaces for $SU(1,1)$.

The spaces $\cD_K$, $K>0$ correspond to one of the discrete series of the representations for
$SU(1,1)$. There is another discrete series of representations. As for the affine \ax group the other discrete series
is obtained by first mapping the matrix in  \eqref{eq:su11matrix} to the matrix with complex conjugate entries.

In each of the representation spaces $\cD_K$, $K\geq0$ we define the family of coherent states
as the unitary transforms of the minimal weight vector $|0\rangle_K$. They can be parametrized by point on the
unit disk $\D$ in $\C$.

\begin{defin}[Coherent states for $SU(1,1)$] For each $z\in \D$ we define  the coherent state
$|z\rangle_K$ by the action of
$$
(1-|z|^2)^{-1/2}\left(\begin{array}{cc} 1&z\\\overline z&1 
  \end{array}\right)
$$
on $|0\rangle_K$.
\end{defin}
The coherent state $|z\rangle_K\in\cD_K$ can be described up to a phase by the property
  $$ 
(z a_1^\dagger+a_2)|z\rangle_K=0.
$$
From this equation we deduce the explicit form
$$
|z\rangle_K=(1-|z|^2)^{(K+1)/2}\sum_{j=0}^\infty \sqrt{j+K \choose K} z^j|j\rangle_K.
$$
We see that
$$
\frac{K}{\pi}\int_{\mathbb D} |z\rangle_K\langle z| (1-|z|^2)^{-2}\dd^2z=I_{\cD_K}.
$$
Using these coherent states we can calculate the coherent state transform of $|\zeta\rangle_K$
\begin{eqnarray*}
\phantom{\langle}_K\langle z|\zeta\rangle_K=(1-|z|^2)^{(K+1)/2}(1-|\zeta|^2)^{(K+1)/2}(1-\overline{z}\zeta)^{-K-1}
\end{eqnarray*}
we see that these indeed agree with the maximizers we conjectured in Section~\ref{sec:analyticform} if $K=2\alpha$.

\begin{conjecture}[Generalized Wehrl for $SU(1,1)$]\label{conj:wehrlsu11}
For any unit vector $\psi\in \cD_K$ and any convex function $G:[0,1]\to\R$ we have that
$$
\int_{\mathbb D} G(|\,{}_K\langle z|\psi\rangle_K|^2)(1-|z|^2)^{-2}\dd^2 z
$$
is maximized if and only if $\psi=|\zeta\rangle_K$ (up to an overall phase) for some $\zeta\in {\mathbb D}$.
\end{conjecture}

This conjecture is equivalent to the \ax conjecture for half-integer $\alpha$.


\section{The $SU(1,1)$ quantum channels}   

\label{sec:su(1.1)channels}

If $K$ and $L$ are non-negative integers
we may identify the representation space $\cD_{K+L}$ as a subspace of $\cD_K\otimes\cD_L$.
Indeed, we can map $|0\rangle_{K+L}$ to $|0\rangle_{K}\otimes |0\rangle_{L}$ and then lift to a representation
by
$$
K_+^n|0\rangle_{K+L}\mapsto (K_+\otimes I_{\cD_L}+I_{\cD_K}\otimes K_+)^n |0\rangle_{K}\otimes |0\rangle_{L}.
$$
Let us denote by $P$ the projection from $\cD_K\otimes\cD_L$ onto $\cD_{K+L}$ considered as a subspace of
$\cD_K\otimes\cD_L$.

We then have a covariant quantum channel (completely positive trace preserving map)$$
\Phi^L(\rho)=c P\rho\otimes I_LP
$$
(for an appropriate $c$) from density matrices on $\cD_K$ to density matrices on  $\cD_{K+L}$.
We conjecture that

\begin{conjecture}[Majorization of $\Phi^L$] \label{conj:majorization} For all density matrices $\rho$ on $\cD_K$
the eigenvalues of $\Phi^L(\rho)$ will be majorized by the eigenvalues of
$\Phi^L(|z\rangle_K\langle z|)$.
\end{conjecture}

\begin{thm}
Conjecture~\ref{conj:majorization} implies Conjecture~\ref{conj:wehrlsu11} except for the uniqueness of the maximizers.
\end{thm}
We shall not give the proof of this theorem here, but it goes along the same lines as the proof of Theorem~2.1 in
\cite{LS1}.



\begin{acknowledgments}

Many thanks to Rupert Frank and Antti Haimi for valuable suggestions
about references  and for suggesting the expansion of the inititial version of this paper from
$AX+B$ to \su. Thanks also to John Klauder who helped us understand the \ax group.
This work was supported by the Villum Centre of Excellence for the
Mathematics of Quantum Theory (QMATH) and the ERC Advanced grant
321029.
\end{acknowledgments}


\end{document}